\documentclass{emulateapj}

\usepackage{color} 
\usepackage{amsmath}
\usepackage{graphicx}
\usepackage{subfigure}
\usepackage{longtable}
\usepackage{multirow}
\usepackage{url}
\usepackage{hyperref}
\usepackage{ulem}

\newcommand\T{\rule{0pt}{3ex}}       
\newcommand\M{\rule{0pt}{2.5ex}}       
\newcommand\dd{\hbox{d}}
\newcommand{\E}{\mathbb{E}}
\newcommand{\R}{\mathbb{R}}
\newcommand{\cR}{\mathcal{R}}
\newcommand{\n}{{\rm{N}}}
\newcommand{\bX}{\boldsymbol{X}}
\newcommand{\bY}{\boldsymbol{Y}}
\newcommand{\V}{\mathcal{V}}
\newcommand{\Cov}{\hbox{Cov}}
\newcommand{\Var}{\hbox{Var}}
\newcommand{\MCz}{{\rm MC}$\_${\rm z}}
\newcommand{\MCztwo}{{\rm MC}$\_${\rm z}{\hskip1pt}2}

\shorttitle{Distance Correlation Methods for Large Astrophysical Databases}
\shortauthors{E. Mart\'{\i}nez-G\'omez, M. T. Richards and D. St. P. Richards}

\begin{document}

\title{Distance Correlation Methods for Discovering Associations in Large Astrophysical Databases}


\author{Elizabeth Mart\'{\i}nez-G\'omez\altaffilmark{1}, Mercedes T. Richards\altaffilmark{2,4}, and Donald St. P. Richards\altaffilmark{3,4}} 

\affil{
{$^1$}Department of Statistics, Instituto Tecnol\'ogico Aut\'onomo de M\'exico, Del. \'Alvaro Obreg\'on, 04510, M\'exico D. F., M\'exico \\ elizabeth.martinez@itam.mx \\
{$^2$}Department of Astronomy \& Astrophysics, Pennsylvania State University, University Park, PA 16802, U.S.A. \\ mrichards@astro.psu.edu \\
{$^3$}Department of Statistics, Pennsylvania State University, University Park, PA 16802, U.S.A. \\ richards@stat.psu.edu \\
{$^4$}Institut f\"ur Angewandte Mathematik, Ruprecht-Karls-Universit\"at Heidelberg, Im Neuenheimer Feld 294, 69120 Heidelberg, Germany
}

\bigskip
\bigskip


\begin{abstract}
High-dimensional, large-sample astrophysical databases of galaxy clusters, such as the Chandra Deep Field South COMBO-17 database, provide measurements on many variables for thousands of galaxies and a range of redshifts. Current understanding of galaxy formation and evolution rests sensitively on relationships between different astrophysical variables; hence an ability to detect and verify associations or correlations between variables is important in astrophysical research. In this paper, we apply a recently defined statistical measure called the {\it distance correlation coefficient}, which can be used to identify new associations and correlations between astrophysical variables. The distance correlation coefficient applies to variables of any dimension; can be used to determine smaller sets of variables that provide equivalent astrophysical information; is zero only when variables are independent; and is capable of detecting nonlinear associations that are undetectable by the classical Pearson correlation coefficient. Hence, the distance correlation coefficient provides more information than the Pearson coefficient. We analyze numerous pairs of variables in the COMBO-17 database with the distance correlation method and with the maximal information coefficient.  We show that the Pearson coefficient can be estimated with higher accuracy from the corresponding distance correlation coefficient than from the maximal information coefficient. For given values of the Pearson coefficient, the distance correlation method has a greater ability than the maximal information coefficient to resolve astrophysical data into highly concentrated horseshoe- or V-shapes, which enhances classification and pattern identification. These results are observed over a range of redshifts beyond the local universe and for galaxies from elliptical to spiral.
\end{abstract}


\section{Introduction}

As we probe deeper into the observable universe in search of a clearer understanding of galaxy formation and evolution, it becomes increasingly more difficult to distinguish between different galaxy types at these higher redshifts, and hence there is a need for techniques that can be used to detect and verify associations and correlations between galaxy properties.  Several high-dimensional, large-sample astrophysical databases have been studied towards this end.  

Many galaxy cluster studies have concentrated on low redshift observations and newer studies have moved beyond the local universe to $z \sim 5$.  One such study included the Chandra Deep Field South region of the sky and resulted in the COMBO-17 (``Classifying Objects by Medium-Band Observations in 17 filters'') database.  This publicly-available catalog was developed by  \citet{wol03a,wol03b,wol04} and includes 63,501 galaxies, stars, quasars, and unclassified objects, with brightness measurements in 17 passbands over the wavelength range 3500 -- 9300 \AA.  The COMBO-17 catalog can be used to perform a statistical investigation of the relationships between the many measured properties associated with galaxies, and the results will be of much interest to both the astrophysics and the statistics communities.

Statistical studies of associations in astrophysical databases have generally been based on the {\it Pearson correlation coefficient}, the classical measure of {\it linear} relationships between two variables \citep{pea1895}.  In the case of the COMBO-17 database, \cite{ric06} outlined a multivariate statistical analysis based on the Pearson correlation coefficients for variables in the catalog; this analysis confirmed correlations between sets of variables that were known to astronomers to be highly correlated.  \citet{izen08} constructed plots of pairwise canonical variables from the COMBO-17 galaxy data and also confirmed similar high correlations between some variables in the catalog.  These studies suggest that Pearson correlation coefficients can be used to identify sets of variables in this database that are highly correlated.  

It is well known that some astrophysical variables have {\it nonlinear} relationships.  Therefore, we need a statistical measure that can detect nonlinear relationships between variables in astrophysical databases.  Since the Pearson correlation coefficient generally cannot detect nonlinear associations and often is zero for dependent variables, \citet{sze07,sze09,sze12,sze13} introduced a new measure, called the {\it distance correlation coefficient}, to address the shortcomings of the Pearson coefficient.  

The distance correlation coefficient has the advantage of being applicable to random variables of any dimension, rather than to two-dimensional variables only, and it has been used to detect nonlinear associations that are undetectable by the Pearson correlation coefficient \citep{sze09}.  Moreover, unlike the Pearson coefficient, the distance correlation coefficient is zero if and only if the variables are independent.  

Hence, the distance correlation coefficient provides more information than the Pearson coefficient, and the number of references to the distance correlation method has increased rapidly across a wide variety of fields, including: machine learning \citep{sri11,sej13}, wind-generation of electrical power \citep{due13}, time series analysis of Earth's ionosphere \citep{gro12}, climate change projections \citep{rac12}, and nuclear chemistry \citep{zho12}.  

In this paper, we apply the distance correlation method to variables in the COMBO-17 database.  Specifically, we compare the distance correlation between pairs of variables with the corresponding Pearson correlation coefficient and also with corresponding values of another statistical measure called the maximal information coefficient \citep{res11}.  The primary aim of this work is to establish the distance correlation measure as superior over alternative methods of discovering associations and correlations between variables in large astrophysical databases.  

In \S\ref{measures-of-association}, we define the distance correlation coefficient and the maximal information coefficient.  In \S\ref{combo17}, we describe the COMBO-17 dataset.  In \S\ref{application}, we describe how the distance correlation and maximal information coefficient measures were applied to the data.  The results and discussion are given in \S\ref{results} and the conclusions are provided in \S\ref{conclusions}.  This work represents the first application of the distance correlation method to astrophysical data.

\section{Measures of association}
\label{measures-of-association}

Two sets of random variables are called {\it independent} if any information provided about the observed values of one set of variables does not affect the conditional probability distribution of the other set.  By convention, a measure of dependence between the two sets of random variables is identically zero if the two sets are independent; hence, such a measure is also called a {\it measure of association}. 

Among the many measures of association between random variables, the most famous and enduring is the Pearson correlation coefficient \citep{pea1895}.  Other measures of association have been developed since then, some of which are similar in approach to Pearson's notion of correlation.  This class of alternative measures includes R\'enyi's maximal correlation \citep{ren59}, rank correlation \citep{spe04,ken38}, and maximal linear correlation \citep{hir35}.  

In recent years, there have appeared several new approaches to measuring association.  These include the maximal information coefficient \citep{res11} and distance correlation coefficient \citep{sze07,sze09,sze12,sze13}.  We describe below the Pearson correlation coefficient and these two new measures in detail.  

Throughout the paper, we assume that all random variables have finite means and variances.

\subsection{The Pearson Correlation Coefficient}
\label{pearson}

Let $X$ and $Y$ be scalar random variables.  We denote the {\it mean} or {\it expectation} of $X$ by $\E(X)$.  The {\it variance} of $X$ is given by $\Var(X) = \E(X^2) - (\E(X))^2$ and the {\it covariance} between $X$ and $Y$ is $\Cov(X,Y) = \E(XY) - \E(X)\E(Y)$.  If $X$ and $Y$ are independent then $\E(XY) = \E(X)\E(Y)$ and therefore $\Cov(X,Y) = 0$.  

The {\it Pearson correlation coefficient} between $X$ and $Y$ is defined to be 
$$
\frac{\Cov(X,Y)}{\sqrt{\Var(X)} \cdot \sqrt{\Var(Y)}} \, .
$$
This coefficient measures the strength of any {\it linear} relationship between the variables since the coefficient equals $\pm 1$ if $X$ and $Y$ are linearly related.  

This correlation coefficient also satisfies many properties that are desirable of measures of association \citep{sch81}.  In particular, if $X$ and $Y$ are independent then it follows that this correlation coefficient equals $0$.  However, the converse is not valid because the coefficient is zero for many dependent variables that satisfy nonlinear relationships.  Consequently, the Pearson coefficient generally is incapable of detecting nonlinear associations between the variables $X$ and $Y$.  

For a random sample $\{(x_i,y_i), i=1,\ldots,\n\}$ drawn from the joint distribution of $(X,Y)$ the {\it empirical}, or {\it sample}, Pearson correlation coefficient is well known to be given by the explicit formula, 
\begin{equation}
\label{eq:empirical-Pearson}
\frac{\sum_{i=1}^\n (x_i-\bar{x})(y_i-\bar{y})}{\sqrt{\sum_{i=1}^\n (x_i-\bar{x})^2} \cdot \sqrt{\sum_{i=1}^\n (y_i-\bar{y})^2}} \, ,
\end{equation}
where $\bar{x} = \n^{-1}\sum_{i=1}^\n x_i$ and $\bar{y} = \n^{-1}\sum_{i=1}^\n y_i$ are the respective sample means.  

\subsection{The Maximal Information Coefficient}
\label{mic}

\citet{res11} recently proposed another measure, the {\it maximal information coefficient} (MIC), to assess the strength of any linear or nonlinear association between two variables.  The MIC is designed mainly for large data sets and is based on Shannon's {\it mutual information criterion} and the related concept of {\it entropy} \citep{sha49,cov91}.  

Let $X$ be a random variable with probability density function $f_1(x)$.  Then the entropy of $X$ is 
$$
H(X) = - \E \log_2 \, f_1(X).
$$
It is well known that entropy is a measure of uncertainty: the higher the entropy, the greater the uncertainty about $X$.  Also, entropy satisfies the property that $H(X) \ge 0$.  

The above definition of entropy extends to a pair of random variables $(X,Y)$ with joint probability density function $f(x,y)$.  We define the joint entropy of $(X,Y)$ to be 
$$
H(X,Y) = - \E \log_2 \, f(X,Y).
$$
Let $f_1(x)$ and $f_2(y)$ denote the marginal probability density functions of $X$ and $Y$, respectively.  Since the function $f(x,y)/f_2(y)$ is the conditional density function of $X$ given $Y$, we also define the {\it conditional entropy} of $X$ given $Y$ to be 
$$
H(X|Y) = - \E \log_2 \, \frac{f(X,Y)}{f_2(Y)}.
$$

The {\it mutual information} $I(X,Y)$ is defined to be 
\begin{equation}
\label{eq:mutual-information}
I(X,Y) = \E \log_2 \frac{f(X,Y)}{f_1(X)f_2(Y)}.
\end{equation}
Note that entropy and mutual information are related through the calculation,  
\begin{equation}
\label{eq:entropy-diff}
\begin{split}
I(X,Y) &= \E \log_2 \left(\frac{1}{f_1(X)} \cdot \frac{f(X,Y)}{f_2(Y)}\right) \\
&= \E\left(-\log_2 f_1(X) + \log_2 \frac{f(X,Y)}{f_2(Y)}\right) \\
&= -\E\log_2 f_1(X) + \E\log_2 \frac{f(X,Y)}{f_2(Y)} \\
&= H(X)-H(X|Y).
\end{split}
\end{equation}
Since Eq. (\ref{eq:mutual-information}) is symmetric in $X$ and $Y$ then it follows that $I(X,Y) = I(Y,X)$.  Hence, it follows from (\ref{eq:entropy-diff}) that the difference in uncertainty about $X$ given knowledge of $Y$ equals the difference in uncertainty about $Y$ given knowledge of $X$.

Turning to the mutual information criterion of \citet{res11}, suppose that we collect a random sample, 
$$
D = \{(x_i,y_i),i=1,2,\ldots,\n\},
$$
drawn from the random variable $(X,Y)$.  We decompose the range of $x$-coordinates of the data into non-overlapping intervals and we also decompose the range of $y$-coordinates of the data into non-overlapping intervals.  These intervals together give rise to a rectangular grid of nonoverlapping bins on the scatterplot of the data.  

Denote by $R$ and $C$ the total number of row and colulmn intervals, respectively.  For each point $(x,y)$ in the $(r,c)$th rectangular bin, the joint probability density function $f(x,y)$ is estimated by $\widehat{f}\,(r,c)$, the proportion of the sample that falls in the $(r,c)$th rectangular bin.  For $x$ in the $r$th row interval, the marginal density function $f_1(x)$ is estimated by 
$$
\widehat{f}_1(r) = \sum_{c=1}^C \widehat{f}\,(r,c),
$$
This shows that $\widehat{f}_1(r)$ represents the proportion of all $\{x_i: i=1,\ldots,\n\}$ that falls in the $r$th row interval.  Similarly, for $y$ in the $c$th column interval, the marginal density function $f_2(y)$ is estimated by 
$$
\widehat{f}_2(c) = \sum_{r=1}^R \widehat{f}\,(r,c),
$$
So, $\widehat{f}_2(c)$ represents the proportion of all $\{y_i: i=1,\ldots,\n\}$ that falls in the $c$th column interval.  

Then the mutual information in Eq. (\ref{eq:mutual-information}) is estimated by the sum, 
\begin{equation}
\label{naive-MIE}
\widehat{I}_{R,C}(D) = \sum_{r=1}^R \sum_{c=1}^C \widehat{f}\,(r,c) \log_2 \frac{\widehat{f}\,(r,c)}{\widehat{f}_1(r)\widehat{f}_2(c)},
\end{equation}
where the sum is taken over all row intervals $r$ and all column intervals $c$.  The estimator $\widehat{I}_{R,C}(D)$ in Eq. (\ref{naive-MIE}) is called the {\it naive mutual information estimate}, and it can be shown that $\widehat{I}_{R,C}(D) \le 1$.  

The value of $\widehat{I}_{R,C}(D)$ clearly depends on $R$ and $C$ and on the choice of intervals.  \citet{res11} therefore define the {\it maximal information coefficient} (MIC) for the data set $D$ to be 
\begin{equation}
\label{eq:empirical-mic}
MIC(D) = \max_{RC < \n^{0.6}} \frac{\widehat{I}_{R,C}(D)}{\log_2 \min\{R,C\}},
\end{equation}
where the maximum is taken over all rectangular grids, i.e., over all integers $R$ and $C$, such that $RC < \n^{0.6}$.  

The statistic $MIC(D)$ is an estimator of the mutual information $I(X,Y)$ given in Eq. (\ref{eq:mutual-information}).  This statistic exhibits the attractive features of a measure of association in that, as the sample size $\n \to \infty$, $MIC(D)$ converges in probability to $1$ if $X$ and $Y$ satisfy a non-constant non-random relationship; also, $MIC(D)$ converges in probability to $0$ if and only if $X$ and $Y$ are independent.  However, some drawbacks of this statistic have been noted by \citet{sim12} and \citet{kin13}.  

We note that, unlike the empirical Pearson correlation coefficient, there does not exist an explicit formula for the empirical MIC; the maximization in Eq. (\ref{eq:empirical-mic}) must be calculated numerically.

\subsection{The Distance Correlation Coefficient}

The distance correlation measure is based on the Fourier transform, or characteristic function, of sets of random variables and the related characterization of independence \citep{sze07}.  

Let $p$ be a positive integer and $X = (X_1,\ldots,X_p) \in \R^p$ be a random vector.  For a vector $s = (s_1,\ldots,s_p) \in \R^p$, the norm $\|s\| = (s_1^2+\cdots+s_p^2)^{1/2}$ denotes the standard Euclidean norm on $\R^p$.  Further, we denote by $\langle s,X\rangle = s_1X_1+\cdots+s_pX_p$ the standard inner product between $s$ and $X$.  

We also consider a positive integer $q$, a vector $t \in \R^q$, and a random vector $Y \in \R^q$ which is associated with $X$.  The Euclidean norm $\|t\|$ and the inner product $\langle t,Y\rangle$ on $\R^q$ are defined similar to the foregoing.  

The {\it joint characteristic function} of the pair of random vectors $(X,Y)$ is 
$$
\phi_{X,Y}(s,t) = \E \exp\left[\sqrt{-1}\langle s,X\rangle+\sqrt{-1}\langle t,Y\rangle\right].
$$
The {\it marginal characteristic functions} of $X$ and $Y$ are 
$$
\phi_X(s) =  \phi_{X,Y}(s,0) = \E \exp\left[\sqrt{-1}\langle s,X\rangle\right],
$$
and 
$$
\phi_Y(t) = \phi_{X,Y}(0,t) = \E \exp\left[\sqrt{-1}\langle t,Y\rangle\right],
$$
respectively.  It is well known that $X$ and $Y$ are mutually independent if and only if $\phi_{X,Y}(s,t) = \phi_X(s)\phi_Y(t)$ for all $s \in \R^p$ and $t \in \R^q$.  

\citealt{sze07} defined the {\it distance covariance} between the random vectors $X$ and $Y$ as the nonnegative number $\mathcal{V}(X,Y)$ defined by
\begin{equation}
\label{dcov}
\begin{aligned}
\V^2(&X,Y) \\
= \, &\frac{1}{c_{p}c_{q}}\int_{\R^q}\int_{\R^p}\frac{|\phi_{X,Y}(s,t)-\phi_{X}(s)\phi_{Y}(t)|^{2}}{\|s\|^{p+1} \, \|t\|^{q+1}} \, \dd s\, \dd t,
\end{aligned}
\end{equation}
where $|v|$ denotes the modulus of the complex number $v$ and 
$$
c_p = \frac{\pi^{(p+1)/2}}{\Gamma\big((p+1)/2\big)}.
$$
The {\it distance correlation} between $X$ and $Y$ is 
\begin{equation}
\label{dcorr}
\cR(X,Y) = \frac{\V(X,Y)}{\sqrt{\V(X,X)} \cdot \sqrt{\V(Y,Y)}}
\end{equation}
if both $\V(X,X)$ and $\V(Y,Y)$ are positive, and defined to be $0$ otherwise.  \citet{sze07} showed that $0 \le \cR(X,Y) \le 1$.  Further, since $X$ and $Y$ are independent if and only if  $\phi_{X,Y}(s,t) = \phi_X(s)\phi_Y(t)$ for all $s$ and $t$, then it follows from Eqs. (\ref{dcov}) and (\ref{dcorr}) that $\cR(X,Y) = 0$ if and only if $X$ and $Y$ are independent.  This is a clear advantage of the distance correlation coefficient over the Pearson correlation coefficient.  Another advantage of the distance correlation $\cR(X,Y)$ over other concepts of correlation is that it is defined for vectors $X$ and $Y$ of arbitrary dimension.  

Despite the higher-dimensional context, \citet{sze07} derived from (\ref{dcov}) and (\ref{dcorr}) a remarkably explicit formula for the corresponding empirical distance correlation:  For a random sample $(\bX,\bY) = \{(X_1,Y_1),\ldots,(X_\n,Y_\n)\}$ from the joint distribution of $(X,Y)$, define for $k=1,\ldots,\n$ and $l=1,\ldots,\n$, 
\begin{equation*}
\begin{aligned}
a_{kl} =& \, \|X_{k}-X_{l}\|_{p}, \\
\bar{a}_{k\cdot} =& \,\frac{1}{\n}\sum_{l=1}^{\n} a_{kl}, \quad \bar{a}_{\cdot l}=\frac{1}{\n}\sum_{k=1}^{\n} a_{kl}, \\
\bar{a}_{\cdot\cdot} =& \, \frac{1}{\n^{2}} \sum_{k,l=1}^{\n} a_{kl},
\end{aligned}
\end{equation*}
and
$$
A_{kl}=a_{kl}-\bar{a}_{k\cdot}-\bar{a}_{\cdot l}+\bar{a}_{\cdot \cdot} \, .
$$
Similarly, define 
\begin{equation*}
\begin{aligned}
b_{kl} =& \, \|Y_{k}-Y_{l}\|_{q}, \\
\bar{b}_{k\cdot} =& \, \frac{1}{\n}\sum_{l=1}^{\n}b_{kl}, \quad \bar{b}_{\cdot l} = 
\frac{1}{\n}\sum_{k=1}^{\n} b_{kl}, \\
\bar{b}_{\cdot \cdot} =& \, \frac{1}{\n^{2}} \sum_{k,l=1}^{\n} b_{kl},
\end{aligned}
\end{equation*}
and
$$
B_{kl} = b_{kl}-\bar{b}_{k\cdot}-\bar{b}_{\cdot l}+\bar{b}_{\cdot \cdot} \, .
$$

The {\it empirical distance covariance} for the random sample $(\bX,\bY)$ is defined to be 
$$
\V_\n(\bX,\bY) = \frac{1}{\n}\Bigg(\sum_{k,l=1}^{\n} A_{kl}B_{kl}\Bigg)^{1/2}.
$$
The {\it empirical distance variance} for the data $\bX = \{X_1,\ldots,X_\n\}$ is defined to be 
$$
\V_\n(\bX) = \frac{1}{\n}\Bigg(\sum_{k,l=1}^{\n}A_{kl}^2\Bigg)^{1/2} \, ;
$$
similarly, the empirical distance variance for the data $\bY = \{Y_1,\ldots,Y_\n\}$ is defined to be 
$$
\V_\n(\bY) = \frac{1}{\n}\Bigg(\sum_{k,l=1}^{\n} B_{kl}^2\Bigg)^{1/2}.
$$

The {\it empirical distance correlation} for the observed data $(\bX,\bY)$ is defined as 
\begin{equation}
\label{eq:empirical-dcor}
\cR_\n(\bX,\bY) = \frac{\V_\n(\bX,\bY)}{\sqrt{\V_\n(\bX)} \cdot \sqrt{\V_\n(\bY)}}
\end{equation}
if both $\V_\n(\bX)$ and $\V_\n(\bY)$ are positive; otherwise, $\cR_\n(\bX,\bY)$ is defined to be 0.  

We remark that the empirical distance correlation coefficient defined in (\ref{eq:empirical-dcor}) has the significant advantage of exhibiting higher {\it statistical power} than the Pearson coefficient and MIC \citep{sze09,sim12,kin13}.  In summary, the distance correlation is more general and more powerful than the Pearson and MIC correlation measures, and these coefficients will now be compared through application to the COMBO-17 astrophysical dataset.

\section{Description of the COMBO-17 Catalog}
\label{combo17}

The COMBO-17 project was carried out largely to study the evolution of galaxies and their associated dark matter halos at $z \le 1$ and the evolution of quasars at $1 \le z \le 5$.  This spectrophotometric survey covers 1 square degree of sky in 17 filters, over a range of wavelengths from $3500- 9300$ \AA, and over five regions of the sky: the Chandra Deep Field South (CDFS), Abell 901, S 11, South Galactic Pole, and Abell 226 fields.  All observations were collected with the Wide Field Imager $0.5^{\circ}\times 0.5^{\circ}$ camera on the MPG/ESO 2.2 m telescope at the European Southern Observatory at La Silla, Chile \citep{wol03a,wol03b,wol04}.  

The CDFS portion of the survey resulted in the detection of 63,501 astronomical objects including over 50,000 galaxies, thousands of stars, hundreds of quasars, and other unclassified objects.  Only $\sim$25,000 of the galaxies have precise photometric redshifts. This data acquisition permits the spectral classification of stars, galaxies, and quasars, as well as the determination of spectral energy distributions and redshifts for galaxies and quasars. The classification is mostly reliable for magnitudes $R \le 24$, while the selection of stars is complete to $R\sim 23$, and deeper for M stars \citep{wol04}.  A catalog calibration update was released by \citet{wol08}.

The COMBO-17 catalog lists identifiers, positions, magnitudes, morphologies, object classification, and redshift information.  It also provides rest-frame luminosities in Johnson, SDSS, and Bessel passbands, and estimated errors.  The COMBO-17 data are available at the website \url{http://www.mpia.de/COMBO/combo_index.html}.  A detailed description of the column entries in the FITS and ASCII versions of the catalog are provided by \citet{wol04} and also on that website.  

The COMBO-17 catalog has been applied to many aspects of cosmology, including galaxy evolution (e.g., \citealt{wol03a,bel04}), the evolution of faint AGN for $1 \le z \le 5$ \citep{wol03b}, weak lensing studies (e.g., \citealt{gra02,kle05}), and star formation in supercluster galaxies \citep{gra04}.

\section{Application of the Distance Correlation Measure to the COMBO-17 Database}
\label{application}

For the application to astrophysics, we concentrated on the galaxies in the COMBO-17 catalog.  We selected 33 variables from the list given in Table 3 of \citet{wol04}.  Of these variables, 5 contain general information about each object, 4 correspond to classification results, 3 are total restframe luminosities, and 21 are observed seeing-adaptive aperture fluxes in observing runs D, E, and F.  Table \ref{table1} lists the variables that were selected for our analysis and their definitions.  

\begin{table}[t]
\caption{Description of Selected COMBO-17 Variables}
\label{table1}
\centering
\begin{tabular}{l|l}
\hline\hline 
\multicolumn{2}{l}{General Information} \M \\ [0.5ex]\hline \T
Rmag    & Total $R$-band magnitude\\
mu\_max & Central surface brightness\\ 
MajAxis & Major axis\\ 
MinAxis & Minor axis\\ 
PA      & Position angle\\ [0.5ex] 
\hline  
\multicolumn{2}{l}{Classification Results} \M  \\ [0.5ex] \hline \T
MC\_z     & Mean redshift in distribution $p(z)$\\ 
MC\_z2    & Alternative redshift if distribution $p(z)$ is bimodal\\ 
MC\_z\_ml & Peak redshift in distribution\\ 
dl        & Luminosity distance of MC\_z\\ [0.7ex] 
\hline  
\multicolumn{2}{l}{Total Object Restframe Luminosities} \M \\ [0.5ex]\hline \T
BjMag & $M_{\rm{abs,gal}}$ in Johnson $B$ ($z \approx [0.0,1.1]$)\\ 
rsMag & $M_{\rm{abs,gal}}$ in SDSS $r$ ($z \approx [0.0,0.5]$)\\ 
S280Mag & $M_{\rm{abs,gal}}$ in 280/40 ($z \approx [0.25,1.3]$)\\ [0.7ex] 
\hline  
\multicolumn{2}{l}{Observed Seeing-Adaptive Aperture Fluxes} \M \\ [0.5ex] \hline \T
W420F\_E & Photon flux in filter 420 in run E\\
W462F\_E & Photon flux in filter 462 in run E\\
W485F\_D & Photon flux in filter 485 in run D\\
W518F\_E & Photon flux in filter 518 in run E\\
W571F\_D & Photon flux in filter 571 in run D\\
W571F\_E & Photon flux in filter 571 in run E\\
W604F\_E & Photon flux in filter 604 in run E\\
W646F\_D & Photon flux in filter 646 in run D\\
W696F\_E & Photon flux in filter 696 in run E\\
W753F\_E & Photon flux in filter 753 in run E\\
W815F\_E & Photon flux in filter 815 in run E\\
W856F\_D & Photon flux in filter 856 in run D\\
W914F\_D & Photon flux in filter 914 in run D\\
W914F\_E & Photon flux in filter 914 in run E\\
UF\_F    & Photon flux in filter U in run F\\
BF\_D    & Photon flux in filter B in run D\\
BF\_F    & Photon flux in filter B in run F\\
VF\_D    & Photon flux in filter V in run D\\
RF\_D    & Photon flux in filter R in run D\\
RF\_E    & Photon flux in filter R in run E\\
RF\_F    & Photon flux in filter R in run F\\   [0.7ex]
\hline
\end{tabular}
\end{table}

In our analysis, we used only flux values that were listed as positive, and we did not consider the estimated errors in the variables.  In addition, we included only galaxies with complete measurements of all 33 variables; hence, galaxies with incomplete data were omitted from our study.  As a consequence of this selection process, our data set contained only 14 galaxies in the range $2 \le z < 3$, so we excluded those galaxies from further analysis.  The final data set consists of 15,352 galaxies over a redshift range $0 \le z < 2$.

\begin{figure}[!t]
\centering
\includegraphics[scale=0.5]{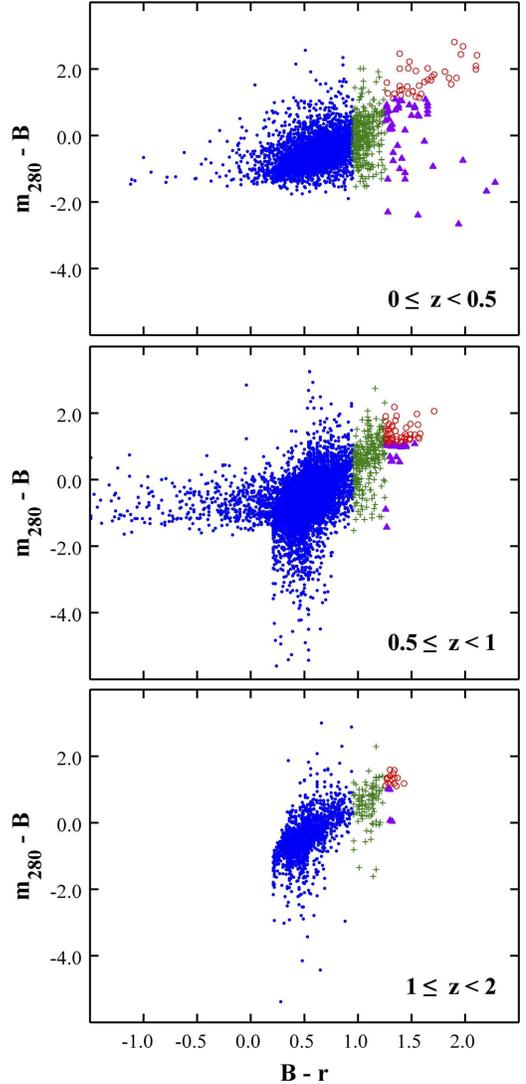}
\caption{Galaxy types based on their $m_{280}-B$ and $B-r$ colors for $0\leq z<0.5$ (upper), $0.5\leq z<1$ (middle), and $1\leq z<2$ (lower):  Type 1 (open circles, red), Type 2 (triangles, purple), Type 3 (plus signs, green), and Type 4 (solid circles, blue).}
\label{fig1}
\end{figure}

\begin{table}[h]
\caption{Galaxy Types and Selected Magnitude Ranges}
\label{table2}
\centering
\begin{tabular}{l|l|c}
\hline\hline \M
Galaxy & Kinney et al. & Magnitude Range based on \\
Type    & 1996 Template     & Fig. 2 of Wolf et al. (2003a)   \\[0.5ex]\hline \T
Type 1  & E - Sa  & {$B-r > 1.25$} ~and~ {$m_{280}-B \geq 1.1$} \\
Type 2  & Sa - Sbc  & {$B-r > 1.25$} ~and~ {$m_{280}-B < 1.1$} \\
Type 3  & Sbc - SB6 & {$0.95 < B-r \leq 1.25$} \\
Type 4  & SB6 - SB1  & {$B-r \leq 0.95$} \\ [0.5ex]
\hline
\end{tabular}
\end{table}

The data were partitioned into four galaxy types and three redshift ranges.  Table \ref{table2} shows how the data were subdivided by galaxy type according to their $m_{280}-B$ and $B-r$ colors; this scheme is similar to the magnitude ranges defined in Figure 2 of \citet{wol03a}, which is based on the galaxy classification template of \citet{kin96} for elliptical and spiral galaxies.  \citet{wol03a} defined these four galaxy types over the redshift range $0.2<z<1.2$, and we extended their scheme for redshifts up to $z=2$.  We also subdivided the data into three redshift bins, as shown in Table \ref{table3}, and we analyzed the individual and combined redshift groups.  Figure \ref{fig1} illustrates the galaxy types for each redshift range based on their $m_{280}-B$ and $B-r$ colors.  

\begin{table}[t]
\caption{Galaxy Analysis Scheme}
\label{table3}
\centering
\begin{tabular}{c|c|c|c|c|c}
\hline\hline  \M
 & \multicolumn{4}{c}{Number of Galaxies} \M \\ [0.5ex] \hline \T
Redshift           &  Type 1  & Type 2 & Type 3  & Type 4 & Total \\ [0.5ex]\hline \T
~~$0 \le z < 0.5$  &    ~~38  &   ~~45 &   ~328  &  ~3254 & ~3665 \\
$0.5 \le z < 1$~~  &    ~~50  &   ~~19 &   ~277  &  ~9284 & ~9630 \\
~~$1 \le z < 2$~~  &    ~~16  &   ~~~4 &   ~109  &  ~1928 & ~2057 \\ [0.5ex]
\hline
Total              &    ~104  &   ~~68 &   ~714  &  14466 & 15352 \\ [0.5ex]
\hline
\end{tabular}
\vspace{5pt}
\end{table}

For the set of 33 variables, there are $(33 \times 32)/2 = 528$ possible pairs of variables.  For each pair, we calculated the empirical Pearson correlation coefficient, MIC, and distance correlation coefficient for each galaxy type and redshift range.  We calculated the empirical Pearson coefficients in Eq. (\ref{eq:empirical-Pearson}) and the empirical MIC scores in Eq. (\ref{eq:empirical-mic}) using software provided by \citet{res13}, and we computed the empirical distance correlation coefficients in Eq. (\ref{eq:empirical-dcor}) with the {\it Energy-Statistics} package of \citet{riz13}.

\section{Results and Discussion}
\label{results}

In this section, we describe the main results from the application of the three statistical measures to the COMBO-17 data; illustrate the effectiveness of the analysis in identifying potential outliers in the data; provide possible explanations for the horseshoe- and V-shaped patterns in the scatterplots; examine the associations between some individual pairs of variables; and discuss the application of the analysis to larger databases, such as the Sloan Digital Sky Survey (SDSS).

\subsection{The COMBO-17 Results}

\begin{figure}[h]
\includegraphics[scale=0.6]{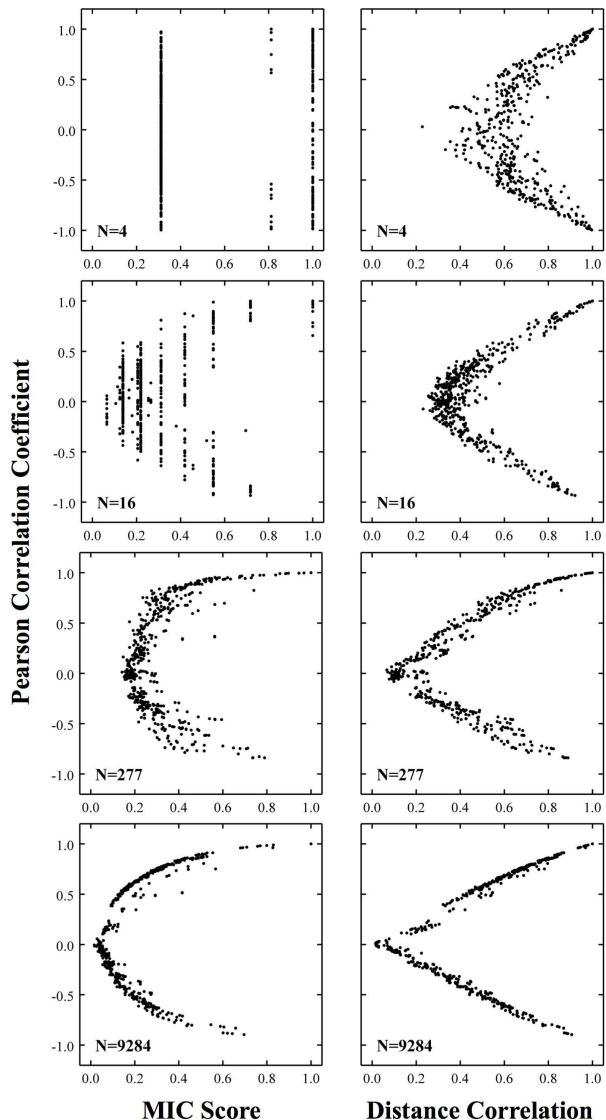}
\caption{Effect of the number of galaxies, $\n$, on the graph of Pearson correlation coefficient vs. MIC score (left frames) compared to the corresponding graph for the distance correlation coefficient (right frames).  These graphs are based on galaxy data, by redshift and type, as indicated in Table \ref{table3}; e.g., the graph with $\n = 277$ is based on the Type 3 galaxies with $0.5 \le z < 1$.}
\label{fig2}
\end{figure}

\begin{figure}[h]
\hspace{-10pt}
 \includegraphics[scale=0.67]{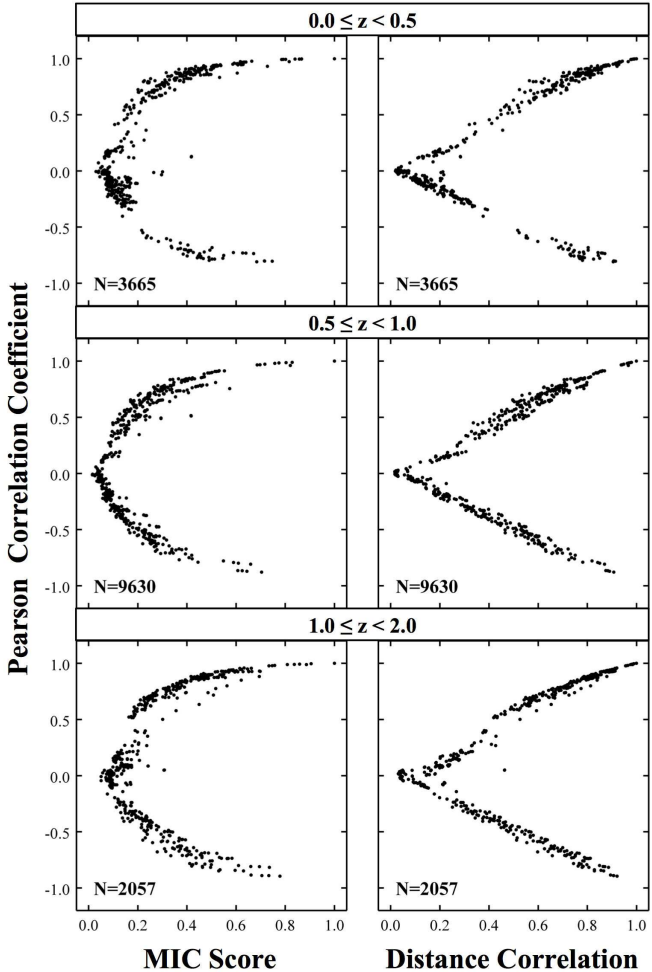}
\caption{The Pearson correlation coefficient versus the MIC score for all galaxies over three redshift ranges, from top to bottom: $0 \le z < 0.5$, $0.5 \le z < 1$, and $1 \le z < 2$.}
\label{fig3}
\vspace{5pt}
\end{figure}

\begin{figure*}[h]
\vspace{-12pt}
\centering
\includegraphics[scale=0.55]{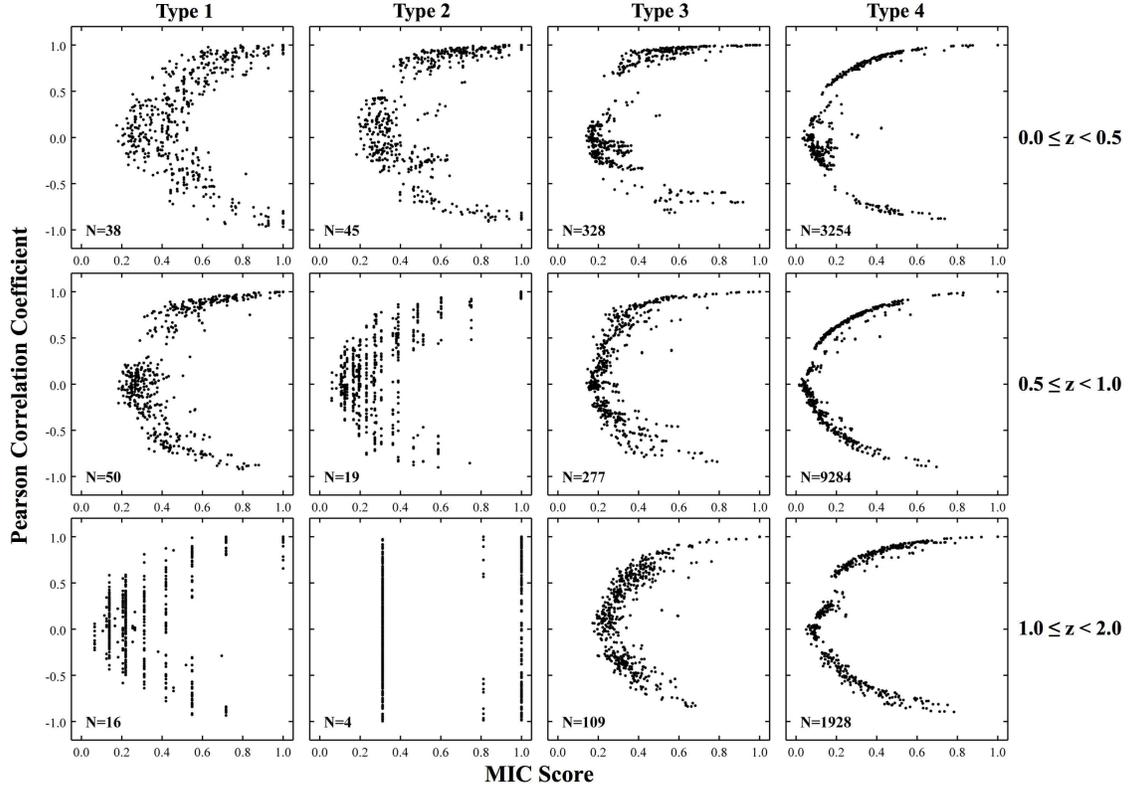}
\caption{The Pearson correlation coefficient versus the MIC score for galaxy types 1 to 4 (columns) and redshift ranges (rows): $0\leq z<0.5$ (upper frames), $0.5\leq z<1$ (middle frames) and $1\leq z<2$ (lower frames).}
\label{fig4}
\end{figure*}

\begin{figure*}[h]
\vspace{-12pt}
\centering
\includegraphics[scale=0.55]{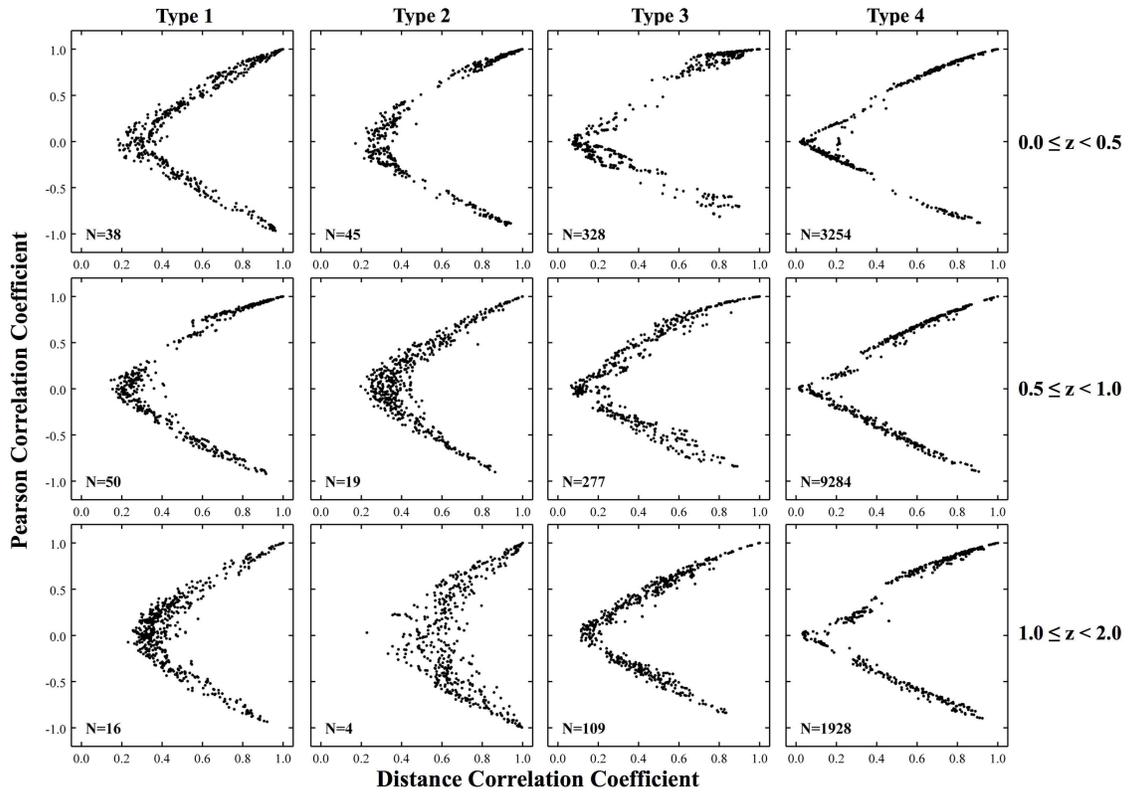}
\caption{The Pearson correlation coefficient versus the distance correlation coefficient for galaxy types 1 to 4 (columns) and redshift ranges (rows): $0\leq z<0.5$ (upper frames), $0.5\leq z<1$ (middle frames) and $1\leq z<2$ (lower frames).}
\label{fig5}
\end{figure*}

The results of our application of distance correlation to the COMBO-17 database are displayed in Figures \ref{fig2} - \ref{fig6} for the four galaxy types and three redshift groups given in Table \ref{table3}.   In these figures, we plot the {\it empirical} correlation coefficients for all 528 pairs of variables based on the list of 33 variables in Table \ref{table1}.  The figures can be interpreted as follows: a low distance correlation coefficient or MIC score suggests a weak statistical relationship between a given pair of variables, while a high distance correlation coefficient or MIC score suggests a strong statistical relationship between the pair.  

Figure \ref{fig2} illustrates the effect of the number of galaxies, $\n$, in the sample on the graph of the Pearson correlation coefficient vs. the MIC score (left frames) compared to the graph of the Pearson coefficient vs. the distance correlation coefficient (right frames).  

\noindent
(1) We see that the overall pattern for both the MIC and distance correlation graphs becomes less diffuse as $\n$ increases, and the relationships become more concentrated and more distinctive for large values of $\n$.  

\noindent
(2) When compared to the distance correlation graphs, the MIC graphs are more influenced by the value of $\n$.  Specifically, the horseshoe-shaped pattern seen for large $\n$ in the MIC graphs breaks down as $\n$ decreases, and leads to sparse values of MIC when $\n$ is very small.  In contrast, the distance correlation graphs display clear V-shaped patterns even for very small sample sizes.  

\noindent
(3) For a given value of $\n$, the relationship between the Pearson and distance correlation coefficients is sharper than the relationship in the case of the MIC score.  This pattern holds even for large values of $\n$.  

Figure \ref{fig3} displays graphs of the Pearson coefficients vs. MIC scores (left frames), and vs. distance correlation coefficients (right frames), for all galaxies over three redshift ranges: $0\leq z< 0.5$, $0.5\leq z<1$, and $1\leq z<2$.

\noindent
(4) The MIC graphs display a horseshoe pattern while the distance correlation graphs display a distinctive V-shaped pattern.  Moreover, the V-shaped pattern for distance correlation is more concentrated than the MIC horseshoe pattern.  Also, the MIC pattern is similar for the three redshift ranges, and the same holds for the distance correlation pattern.

Figures \ref{fig4} and \ref{fig5} provide more detailed versions of Figure \ref{fig3}, where the galaxies have been separated into four types, as listed in Table \ref{table3}.  These subplots display the differences in the scatterplots for the various combinations of galaxy type and redshift.  In Figure \ref{fig4}, the Pearson coefficient is plotted vs. the MIC score for four galaxy types (columns) and three redshift ranges (rows); this figure shows that the horseshoe pattern persists across galaxy types except when $\n$ is low, as noted earlier.  Figure \ref{fig5} shows the behavior of the Pearson vs. distance correlation coefficients over the same grid of galaxy types and redshifts; by contrast, this figure shows that the V-shaped relationship between the Pearson and distance correlation coefficients persists for all values of $\n$, even for low $\n$.  

It is noticeable that the MIC subplots in Figure 4 are less distinctive, especially because these scatterplots are more sensitive to the number of galaxies in the subplot sample.  However, the distance correlation subplots in Figure 5 are much sharper, regardless of the subplot sample size.  Hence it is easier to see that there is general consistency between the latter subplots across the galaxy types and redshift ranges.  This is an advantage of the distance correlation measure.  Since the distance correlation scatterplots are only weakly dependent on sample size, we can see that there are noticeable differences between the V-shaped patterns for the different galaxy types and redshift ranges.

\subsection{Examination of Potential Outliers}

Since the horseshoe-shaped MIC pattern is more diffuse than the V-shaped distance correlation pattern, Figure \ref{fig3} confirms that distance correlation is a stronger measure of association than MIC.  Consequently, the distance correlation measure is more effective than MIC in identifying pairs of variables that are potential outliers, which can then be investigated in greater detail.

\begin{table}[h]
\caption{Distance Correlation Outlier Pairs of Variables \\ for $1 \le z < 2$ in Figure \ref{fig3} }
\label{table4}
\centering
\begin{tabular}{l|c|c}
\hline\hline  \M
Variables           & Distance Correlation & Pearson Coefficient \M \\ 
[0.5ex] \hline \T
(dl,\, MC$\_$z2)      & 0.46278               & 0.04904 \\
(MC$\_$z2,\, MC$\_$z) & 0.46336               & 0.04948 \\ [0.5ex]
\hline
\end{tabular}
\end{table}

A potential outlier pair of variables is noticeable in the bottom right frame of Figure \ref{fig3}.  As shown in Table \ref{table4}, our calculations reveal that this location in the graph is associated with {\it two} pairs of variables: ({dl},\, {\MCztwo}) and ({\MCztwo},\, {\MCz}), where these variables are defined in Table \ref{table1}.  The two pairs are clearly related to each other since {dl} is associated with {\MCz} through Hubble's Law, and the variable {\MCztwo} appears in both pairs.  Since {\MCztwo} is the alternative redshift if the probability distribution $p(z)$ is bimodal \citep{wol04}, then the distance correlation appears to have detected the bimodal nature of this underlying probability distribution.  

\subsection{Interpretation of Horseshoe- and V-shaped Patterns} 

There are several possible reasons for the horseshoe- and V-shaped patterns seen in Figures \ref{fig2}-\ref{fig5}.  The common thread connecting these explanations is that the patterns appear when high-dimensional data are compressed into two-dimensional space. The literature on this phenomenon is extensive, and numerous references on this topic can be found in \citet{dia08}.  

Horseshoe patterns have been found in a variety of settings.  In archaeology and ecology, these patterns are known as the ``horseshoe effect'' \citep{ken70}; and in correspondence analysis, this phenomenon is known as the ``Guttman effect'' \citep{dia08}.  If the data satisfy certain Gaussian distribution properties then many methods of reducing multidimensional data to two dimensions result in horseshoe-shaped plots.  

These patterns also arise when ``kernel-type'' statistics are used to map high-dimensional data into two-dimensional space \citep{dia08}.  The Pearson coefficient and distance correlation coefficient are of kernel type \citep{sej13}, so the horseshoe- and V-shaped plots for the COMBO-17 data could be due to the manner in which the Pearson and distance correlation coefficients are defined. 
 
Another possibility is that horseshoe patterns could be intrinsic to the COMBO-17 data.  Special data models, such as the ``Kac-Murdoch-Szeg\"o model,'' lead to similar patterns when the data are compressed to two dimensions \citep{dia08}.  It would be a remarkable discovery if the COMBO-17 data were shown to satisfy one of these special data models.

In the case of the COMBO-17 database, the reduction of the high-dimensional data to two-dimensional scatterplots of correlation coefficients represents the type of compression that has given rise to horseshoe shapes in other applications.  Although further investigation is required to explain why the COMBO-17 data are clustered in such distinctive ways, the greater significance of these horseshoe- and V-shaped plots is that they provide a mechanism for isolating potential outliers, which can then be analyzed subsequently in greater detail.

\begin{figure*}[t]
\centering
\hspace{-5pt}
\includegraphics[scale=0.53]{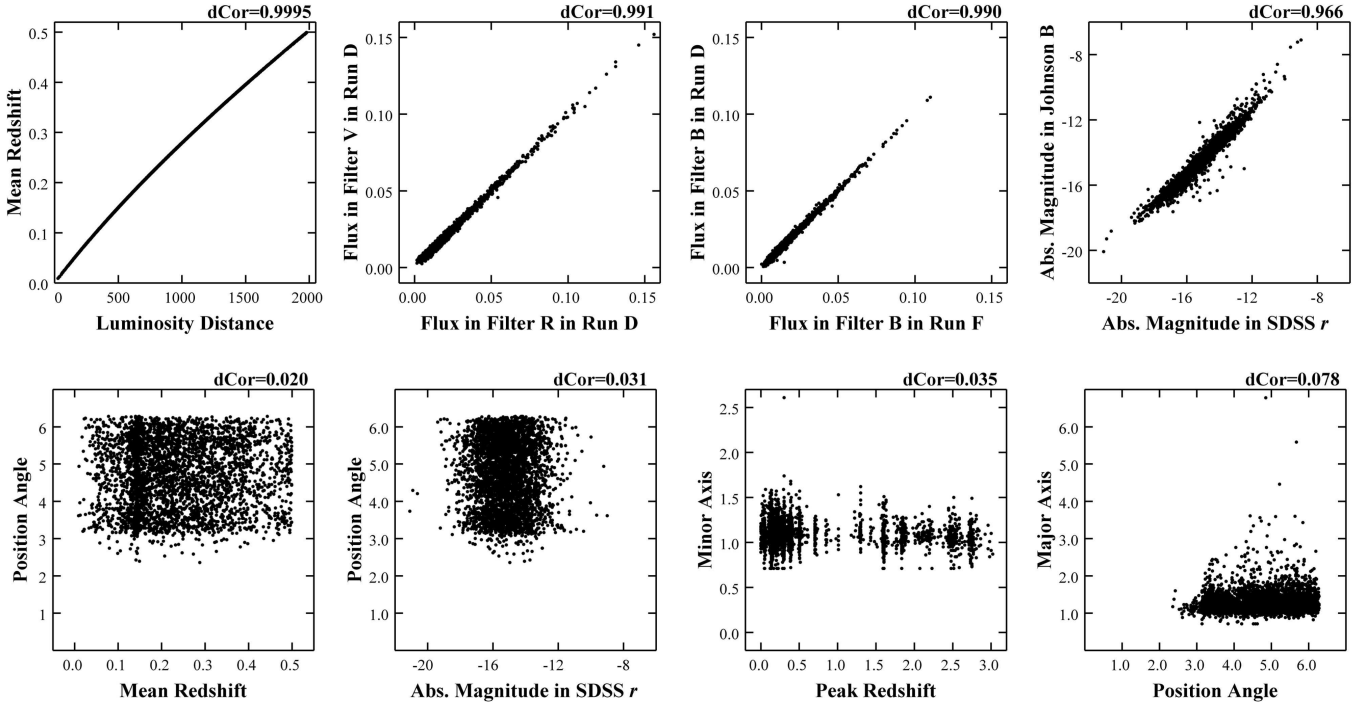}
\caption{Examples of pairs of variables with high distance correlation coefficients (upper frames) and low distance correlation coefficients (lower frames) for all galaxies with redshift $0\leq z<0.5$.}
\label{fig6}
\vspace{16pt}
\end{figure*}

\subsection{Associations between Individual Pairs of Variables}

Figure \ref{fig6} was designed to illustrate the ability of the distance correlation measure to identify associations or correlations that are well known and to provide a consistency check on our analysis.  From the 528 pairs of variables, we selected four pairs with very high distance correlation coefficients (i.e., very strong associations) and four pairs with very low distance correlation coefficients (i.e., very weak associations).  These comparisons were included to illustrate the advantages of the distance correlation method.  
 
Examples of pairs of variables with high distance correlation coefficients include the photon fluxes at selected wavelengths in different observing runs; these fluxes obviously are closely related to each other.  Another example is the pair of variables, redshift and luminosity distance; the high distance correlation coefficient of $0.9995$ confirms their well known association.  

The scatterplots in Figure \ref{fig6} for the pairs with very high distance correlation coefficients may seem, at first glance, to have revealed (nearly) linear relationships between the variables.  However, the apparent linearity of a scatterplot is insufficient to justify the application of Pearson's coefficient as a measure of association.  The Pearson coefficient is applicable only after it is known that a relationship is linear, and it is only then that the Pearson coefficient can be used to measure the strength of a linear association.  

A closer inspection of the upper panels in Figure \ref{fig6} reveals some interesting conclusions.  The Hubble diagram plot of redshift vs. luminosity distance in Frame 1 of Figure \ref{fig6} exhibits a slight curvature over the relatively small redshift range $0 \le z < 0.5$; such a non-linear relationship is in accordance with the latest models for the curvature of the universe.  

An unexpected result is seen in the middle two top frames, corresponding to dCor=0.991 and dCor=0.990.  These scatterplots seem to reveal linear relationships between fluxes in different filters or for the same filter in different observing runs.  However, the thickness of the plots varies with flux range, being thicker for smaller fluxes and thinner for larger fluxes along the horizontal axis; this phenomenon is called {\it heteroskedasticity} in the statistical literature.  In the presence of heteroskedasticity, it is generally the case that the Pearson correlation coefficient and related statistical methods, such as standard linear regression, are not applicable because they assume {\it homoskedasticity}, i.e., constant thickness of the plots for different values along the horizontal axis.  

In the fourth frame, corresponding to dCor=0.966, we can see even greater heteroskedasticity in the scatterplots, indicating that it may be even more unwise to apply the Pearson coefficient in this case to assess the strength of an association between the variables.

Consequently, the upper panels of Figure 6 provide us with stronger basis for believing the results seen in the lower panels of the same figure and also the unanticipated finding of heteroskedasticity. 

Figure \ref{fig6} also shows that certain pairs have distinctly low distance correlation coefficients, and hence weak associations.  For example, the position angle, length of the minor axis of the galaxy, and length of the major axis are found to be weakly associated with redshift since their distance correlation coefficients are negligible.  As well, the position angle is found to be weakly associated with the minor axis or the major axis, again with negligible distance correlation coefficients.

\subsection{Application to Larger Databases}

In the application of distance correlation to larger databases such as the SDSS, the computational and statistical aspects are the main issues.  We discuss these aspects below.   

From the computational perspective, the distance correlation formulas are directly applicable to the SDSS, or to any data set, regardless of the sample size, N, or the number of variables in the set, $p+q$. (Here, the maximum value of $p+q$ is the total number of variables; and in the case of pairs of variables, $p=1$ and $q=1$.)  Equation (\ref{eq:empirical-dcor}) and the preceding formulas for the distance covariance and distance variance show that the empirical distance correlation is straightforward to calculate for any data set.  The calculations may be more time-consuming for larger data sets; however, the computational complexity remains the same.  

From a statistical perspective, the behavior of the distance correlation coefficient when N and $p+q$ are very large depends on the statistical distribution of the data in the catalog.  \citet{sze07} determined the behavior of the distance correlation measure for fixed $p$ and $q$, and for increasing values of N, regardless of the statistical distribution of the parent population from which the data are drawn; this is called an ``asymptotic nonparametric'' result.  

If both N and $p+q$ are very large, and the underlying $(X,Y)$ parent population has a multivariate Gaussian distribution, \citet{due13} derived a comprehensive description of the behavior of the distance correlation measure.  In all remaining cases, in which N and $p+q$ are large and the $(X,Y)$ population is non-Gaussian, the behavior of the distance correlation measure remains generally unknown.  Even in instances in which the $(X,Y)$ population is a mixture of Gaussians, which includes many bimodal and heavy-tailed distributions, the mathematical calculations underlying distance correlation are non-trivial and are still open. 

Nevertheless, as a computational matter, the distance correlation method can be directly applied to large-N and large-($p+q$) data, such as the SDSS catalog.  In that case, the scatterplots may represent superpositions of the horseshoe- or V-shaped patterns of the type seen in the COMBO-17 data.  In fact, Figures \ref{fig2} and \ref{fig3} suggest that the patterns for the COMBO-17 data may already represent superpositions, rather than single, horseshoe- or V-shaped scatterplots.   

Finally, distance correlation procedures can now be implemented automatically inside databases in the same manner as classical statistical methods.  To implement the distance correlation computations in databases, we recommend that researchers use the {\it Energy-Statistics} package of \citet{riz13}.

\section{Conclusions}
\label{conclusions}

There are four aspects to this work: the introduction to the astrophysics community of a new statistical measure of association, called distance correlation; the numerical computations needed to process the data; the application of distance correlation to a large astrophysical database such as the COMBO-17 catalog; and the discovery of a mechanism that accentuates the differences between potential outliers and the remaining data points.  

This paper is the first application in which the Pearson, MIC, and distance correlation measures have been compared directly.  In this paper, we analyzed the associations between pairs of variables, and we have shown that the methods apply equally well to sets of variables of any dimension.  

For the application to the COMBO-17 database, we applied the distance correlation method to 33 variables for a sample of 15,352 galaxies, with redshifts $0 \le z < 2$.  For the corresponding 528 pairs of variables, we compared the Pearson correlation coefficient to the maximal information and distance correlation coefficients.  (1) We found that the relationship between the Pearson and distance correlation coefficients is sharper than the relationship between the Pearson coefficient and the MIC score, regardless of the sample size.  (2) The MIC graphs display a horseshoe pattern while the distance correlation graphs display a more concentrated and distinctive V-shaped pattern; and these patterns remain the same for all redshift ranges.  (3) The MIC graphs are also more influenced by the number of galaxies in the sample; the horseshoe pattern becomes noticeably more diffuse when the number of galaxies is small.  On the other hand, the distance correlation graphs display sharp V-shaped patterns, regardless of sample size.  Hence, the distance correlation is a stronger measure of association than MIC.  (4) The distance correlation is more effective than MIC in identifying pairs of variables that are potential outliers; further, we identified two outlying pairs of variables that are associated with a bimodal distribution of redshifts.  (5) We can also examine the level of association between individual pairs of variables; and we used the distance correlation measure to confirm known associations between pairs of variables that have high distance correlations and identified other pairs that have low distance correlations, and hence are weakly associated.  (6) Our analysis revealed unexpected heteroskedasticity in near-linear relationships between some pairs of variables, which is another advantage of the distance correlation method over the classical Pearson coefficient.  

Our results indicate that the distance correlation measure is superior to alternative methods used to analyze associations between variables in astrophysical databases.  The advantages of the distance correlation method rest in its applicability to groups of random variables of any dimension; its ability to detect nonlinear associations that are undetectable by the Pearson coefficient; its ability to cluster data into V-shaped patterns that can readily be used to identify potential outliers in the data set; and its ability to identify independence between random variables.  Finally, this analysis illustrates the broader applicability of the distance correlation measure to other large databases. 

\acknowledgements
We thank the referee for very helpful comments on the manuscript.  This research was partially supported by National Science Foundation grants AST-0908440 and DMS-1309808.

\qquad


\begin{thebibliography}{}

\bibitem[Bell et al.(2004)]{bel04} 
Bell, E. F., Wolf, C., Meisenheimer, K., Rix, H.-W., Borch, A., Dye, S., Kleinheinrich, M., Wisotzki, L., \& McIntosh, D. H. 2004, ApJ, 608, 752

\bibitem[Cover \& Thomas(1991)]{cov91} 
Cover, T. M., \& Thomas, J. A. 1991, {\it Entropy, Relative Entropy and Mutual Information in Elements of Information } (New York: Wiley)

\bibitem[Diaconis et al.(2008)]{dia08}
Diaconis, P., Goel, S. \& Holmes, S. 2008, Ann. Appl. Stat., 2, 777

\bibitem[Dueck et al.(2013)]{due13}
Dueck, J., Edelmann, D., Gneiting, T., \& Richards, D. 2013, Bernoulli, \href{http://arxiv.org/abs/1210.2482}{in press}

\bibitem[Gray et al.(2002)]{gra02} 
Gray, M. E., Taylor, A. N., Meisenheimer, K., Dye, S., Wolf, C., \&  Thommes, E. 2002, ApJ, 568, 141

\bibitem[Gray et al.(2004)]{gra04} 
Gray, M. E., Wolf, C., Meisenheimer, K., Taylor, A. N., Dye, S., Borch, A. \&  Kleinheinrich, M. 2004, MNRAS, 347, L73

\bibitem[Gromenko et al.(2012)]{gro12}
Gromenko, O., Kokoszka, P., Zhu, L., \& Sojka, J. 2012, Ann. Appl. Statist., 6, 669

\bibitem[Hirschfeld \& Wishart(1935)]{hir35} 
Hirschfeld, H. O., \& Wishart, J. 1935, Math. Proc. Camb. Philos. Soc., 31, 520

\bibitem[Izenman(2008)]{izen08}
Izenman, A. J. 2008, {\it Modern Multivariate Statistical Techniques: Regression, Classification, and Manifold Learning} (New York: Springer), pp. 215--219

\bibitem[Kendall(1970)]{ken70} 
Kendall, D. G. 1970, Phil. Trans. Roy. Soc. London, 269, 125

\bibitem[Kendall(1938)]{ken38} 
Kendall, M. G. 1938, Biometrika, 30, 81

\bibitem[Kinney et al.(1996)]{kin96} 
Kinney, A. L., Calzetti, D., Bohlin, R. C., McQuade, K., Storchi-Bergmann, T., \& Schmitt, H. R. 1996, ApJ, 467, 38

\bibitem[Kinney \& Atwal(2013)]{kin13} 
Kinney, J. B., \& Atwal, G. S. 2013, arXiv 1301.7745v1  

\bibitem[Kleinheinrich et al.(2005)]{kle05} 
Kleinheinrich, M., Rix, H.-W., Schneider, P., Erben, T., Meisenheimer, K., Wolf, C.., \& Schirmer, M. 2005, in IAU Symp. 225, Impact of Gravitational Lensing on Cosmology, ed. Y. Mellier, \& G. Meylan (New York: Cambridge University Press), 249

\bibitem[Pearson(1895)]{pea1895} 
Pearson, K. 1895, Proc. Roy. Soc. London, 58, 240

\bibitem[Racherla et al.(2012)]{rac12}
Racherla, P. N., Shindell, D. T., \& Faluvegi, G. S. 2012, J. Geophys. Res. 117, D20118

\bibitem[R\'enyi(1959)]{ren59} 
R\'enyi, A. 1959, Acta Math. Acad. Sci. Hungarica, 10, 441

\bibitem[Reshef et al.(2011)]{res11} 
Reshef, D. N., Reshef, Y. A., Finucane, H., K., Grossman, S. R., McVean, G., Turnbaugh, P. J., Lander, E. S., Mitzenmacher, M., Sabeti, P. C. 2011, Science, 334, 1518

\bibitem[Reshef \& Reshef(2013)]{res13}
Reshef, D. \& Reshef, Y. 2013, MINE: Maximal Information-based Nonparametric Exploration, \url{http://www.exploredata.net/Usage-instructions}

\bibitem[Richards(2006)]{ric06}
Richards, D. St. P., 2006, Lectures on Multivariate Analysis, \url{http://www.stat.psu.edu/~richards/talks/astro-talks.html}

\bibitem[Rizzo \& Sz\'ekely(2013)]{riz13}
Rizzo, M. L. \& Sz\'ekely, G. J. 2013, Energy Statistics {\sl R}-package, \url{http://cran.us.r-project.org/web/packages/energy/index.html}

\bibitem[Schweizer \& Wolff(1981)]{sch81} 
Schweizer, B. \& Wolff, E. F. 1981, Ann. Statist., 9, 879

\bibitem[Sejdinovic et al.(2013)]{sej13}
Sejdinovic, D., Sriperumbudur, B., Gretton, A. \& Fukumizu, K. 2013, Ann. Statist., 41, 2263

\bibitem[Shannon \& Weaver(1949)]{sha49} 
Shannon, C. E. \& Weaver, W. 1949, {\it The Mathematical Theory of Communication} (University of Illinois Press: Urbana).

\bibitem[Simon \&  Tibshirani(2012)]{sim12}
Simon, N. \& Tibshirani, R. 2012, 
\url{http://www-stat.stanford.edu/~tibs/reshef/comment.pdf}

\bibitem[Spearman(1904)]{spe04} 
Spearman, C. 1904, Am. J. Psychol., 15, 72

\bibitem[Sriperumbudur et al.(2011)]{sri11}
Sriperumbudur, B. K., Fukumizu, K., \& Lanckriet, G. R. G. 2011, J. Mach. Learn. Res., 12, 2389

\bibitem[Sz\'ekely et al.(2007)]{sze07} 
Sz\'ekely, G. J., Rizzo, M. L., \& Bakirov, N. K. 2007, Ann. Statist., 35, 2769

\bibitem[Sz\'ekely et al.(2009)]{sze09} 
Sz\'ekely, G. J., \& Rizzo, M. 2009, Ann. Appl. Statist., 3, 1236

\bibitem[Sz\'ekely et al.(2012)]{sze12}
Sz\'ekely, G. J. \& Rizzo, M. 2012, Statist. \& Probab. Lett., 82, 2278

\bibitem[Sz\'ekely et al.(2013)]{sze13}
Sz\'ekely, G. J. \& Rizzo, M. 2013, J. Multivariate Anal., 117, 193

\bibitem[Wolf et al.(2003a)]{wol03a} 
Wolf, C., Meisenheimer, K., Rix, H.-W., Borch, A., Dye, S., \& Kleinheinrich, M. 2003a, A\&A, 401, 73

\bibitem[Wolf et al.(2003b)]{wol03b} 
Wolf, C., Wisotzki, L., Borch, A., Dye, S., Kleinheinrich, M., \& Meisenheimer, K. 2003b, A\&A, 408, 499

\bibitem[Wolf et al.(2004)]{wol04} 
Wolf, C., Meisenheimer, K., Kleinheinrich, M., Borch, A., Dye, S., Gray, M., Wisotzki, L., Bell, E. F., Rix, H.-W., Cimatti, A., Hasinger, G., Szokoly, G. 2004, A\&A, 421, 913

\bibitem[Wolf et al.(2008)]{wol08}
Wolf, C., Hildebrandt, H., Taylor, E. N., \& Meisenheimer, K. 2008, A\&A, 492, 933

\bibitem[Zhong et al.(2012)]{zho12}
Zhong, J., DiDonato, N., \& Hatcher, P. G. 2012, J. Chemometrics, 26, 150

\end{thebibliography}
\end{document}